\documentclass[pra,twocolumn,showpacs,superscriptaddress]{revtex4}
\usepackage{graphicx}
\usepackage{amssymb}
\usepackage{amsfonts}
\usepackage{amsmath}
\usepackage{lmodern}
\usepackage{mathrsfs}
\usepackage[dvipdfm,colorlinks=true, citecolor=blue, urlcolor=blue, linkcolor=blue ]{hyperref}% add hypertext capabilities
\hypersetup{breaklinks=true}
\begin{document}
\title{Frozen Gaussian quantum discord in photonic crystal cavity array system}
\author{Ying-Qi L\"{u}}
\affiliation{School of Nuclear Science \& Technology, Lanzhou University, Lanzhou 730000, China}
\author{Jun-Hong An}
\email{anjhong@lzu.edu.cn} \affiliation{Center for Interdisciplinary
Studies, Lanzhou University, Lanzhou 730000, China}
\affiliation{Centre for Quantum Technologies, National University of
Singapore, 3 Science Drive 2, Singapore 117543, Singapore}
\author{Xi-Meng Chen}
\affiliation{School of Nuclear Science \& Technology, Lanzhou University, Lanzhou 730000, China}
\author{Hong-Gang Luo}
\affiliation{Center for Interdisciplinary Studies, Lanzhou University, Lanzhou 730000, China}
\author{C. H. Oh}\email{phyohch@nus.edu.sg}\affiliation{Centre for Quantum Technologies, National University of
Singapore, 3 Science Drive 2, Singapore 117543, Singapore}

\begin{abstract}
Protecting quantum correlation from decoherence is one of the crucial issues in quantum information processing. It has been commonly recognized that any initial quantum correlation of a composite system diminishes asymptotically or abruptly to zero under local Markovian decoherence. Here we show that, contrary to this recognition, a noticeable Gaussian quantum discord of a continuous-variable bipartite system can be frozen in the steady state in the non-Markovian dynamics if each of the subsystems forms a localized mode with its local reservoir. The condition for this frozen quantum discord can be reached by appropriately engineering the structure of the reservoirs. The possible realization of our results in a coupled cavity array system formed by a photonic crystal is proposed.
\end{abstract}
\pacs{03.65.Yz, 03.67.Mn, 42.50.Ex}
\maketitle

\section{Introduction}
Quantum correlation plays an essential role in quantum information science. In the early days of quantum information, quantum correlation was characterized by entanglement, which is viewed as the main resource for quantum information processing \cite{Horodecki2009}. It engenders the dramatic speedup of a quantum computer over its classical counterpart. Recently, it was found that entanglement is not the only reason to cause such speedup and that a similar speedup can also be achieved in the so-called deterministic one-qubit quantum computation by use of the zero-entanglement states \cite{Knill1998,Lanyon2008}. It has been attributed to another measure of quantum correlation \cite{Datta2008}, i.e. quantum discord (QD) \cite{Ollivier2001,Henderson2001}. These results indicate that entanglement cannot exhaust quantum correlation and QD characterizes the quantumness of correlations more generally than entanglement.

%Owning to the analytical achievements to quantify QD for two-qubit Bell-diagonal state \cite{Luo2008,Ali2010} and for the bipartite continuous-variable Gaussian state \cite{Giorda2010,Adesso2010},
The study of quantum correlations under decoherence has attracted much attention in recent years, because this study is expected to supply some insight regarding how to overcome the detrimental effects caused by decoherence on quantum correlation. It is found that QD \cite{Luo2008,Ali2010,Giorda2010,Adesso2010} exhibits some peculiar features which are absent for entanglement. First, QD of a two-qubit system under individual decoherence decays to zero in an asymptotical manner \cite{Werlang2009,Fanchini2010,Wang2010,Ferraro2010,Ge2010,Xu2010,Lo1}, which is much different from the sudden death behavior of entanglement in the same setting \cite{Yu2004,Almeida2007}. The experimental \cite{Madsen2012} and theoretical \cite{Vasile2010,Isar2012} works also confirm similar results for the Gaussian QD of continuous-variable systems. Second, QD can be developed transiently from a certain initially classical state under a single local Markovian dissipation channel, both for discrete-variable \cite{Ciccarello2012,Streltsov2011} and continuous-variable \cite{Ciccarello20122} systems. This is unattainable with entanglement. Third, QD under decoherence shows a sudden change from the ``classical decoherence'' regime to the ``quantum decoherence'' regime \cite{Mazzola2010,You2012,Lo2}. In the former regime, the classical correlation decays while QD is frozen to its initial value; in the latter regime, QD starts to decay while classical correlation is frozen. This interesting phenomenon has been observed in optical \cite{Xu2010} and NMR \cite{Auccaise2011} systems.

All of these features indicate that QD is the more robust than entanglement against decoherence. As a result, QD could be more preferred resource in quantum information processing. However, one finds that QD, discussed above, decays exclusively to zero in the long-time limit. To overcome the detrimental effects of decoherence on quantum information processing, it is of course desirable to preserve the initial quantum correlation in the long-time limit.

In this work, we propose a scheme to stabilize QD by appropriately engineering the reservoirs to introduce the non-Markovian effect, an issue actively studied recently \cite{Zhang2012,Ferialdi2012,Huelga2012,Hoeppe2012,Haikka2012}. By studying the correlation dynamics of a continuous-variable bipartite system, we show that a finite Gaussian QD can be frozen in the steady state. The essential physics is the formation of a localized mode in the subsystems and the non-Markovian effect. An experimentally accessible scheme is proposed to observe the frozen QD by using a coupled cavity array system realized especially in a photonic crystal system \cite{Hennessy2007, Notomi2008, Majumdar2012}. The result and its possible experimental realization could be significant in quantum information processing.

\section{Model and dynamics}
Consider two noninteracting harmonic oscillators coupled to two independent reservoirs. The Hamiltonian of each local subsystem is ($\hbar=1$)
\begin{eqnarray}
    \hat{H}^{k}=\omega_k\hat{a}_k^{\dag}\hat{a}_k+\sum_{l}\omega_{kl}\hat{b}_{kl}^{\dag}\hat{b}_{kl}+\sum_{l}(g_{kl}\hat{a}^{\dag}_k\hat{b}_{kl}+ \text{h.c.}),\label{Hamilt}
\end{eqnarray}
where $\hat{a}_k$ and $\hat{b}_{kl}$ ($\hat{a}^{\dag}_k$ and $\hat{b}^{\dag}_{kl}$) are, respectively, the annihilation (creation) operators of the $k$-th harmonic oscillator with frequency $\omega_k$ and its corresponding reservoir. The coupling strength between them is given by $g_{kl}$. The system is highly pertinent to a quantum-optical setting where the system oscillators can describe the quantized optical fields in cavity \cite{Zhou2012} or in circuit \cite{Fink2008} QED, mechanical oscillators in opto-mechanics \cite{Kippenberg2007}, and atomic ensemble under a large-$N$ limit \cite{Hammerer2009}. Currently, most quantum optical experiments are performed at low temperatures and under vacuum condition. Thus, we assume the reservoirs to be at zero temperature in this work.

The exact decoherence dynamics of the system can be derived by Feynman and Vernon's influence-functional theory \cite{Feynman63,An2007}. The reduced density matrix of the system expressed in the coherent-state representation is given by
\begin{eqnarray}\label{rout}
\rho (\boldsymbol{\bar{\alpha}}_{f},\boldsymbol{\alpha }_{f}^{\prime };t)&=&\int d\mu (\boldsymbol{\alpha }_{i})d\mu (\boldsymbol{\alpha }
_{i}^{\prime })\mathcal{J}(\boldsymbol{\bar{\alpha}}_{f},\boldsymbol{\alpha }_{f}^{\prime };t|\boldsymbol{\bar{\alpha}}_{i},\boldsymbol{\alpha }
_{i}^{\prime };0)  \notag \\
&&~~~~~~~~~\times \rho (\boldsymbol{\bar{\alpha}}_{i},\boldsymbol{\alpha }_{i}^{\prime };0).
\end{eqnarray}%
The coherent-state representation is defined as $|\boldsymbol{\alpha}\rangle = \prod_{k = 1}^2\exp (\alpha _ka^{\dagger}_k)|0_k\rangle$, which are the eigenstates of annihilation operators and obey the resolution of identity, $\int d\mu \left( \boldsymbol{\alpha}\right) |\boldsymbol{\alpha }\rangle \langle \boldsymbol{\alpha }| = 1$ with the integration measures defined as $d\mu\left(\boldsymbol{\alpha}\right) = \prod_{k}e^{-\bar{\alpha}_{k}\alpha_{k}}\frac{d\bar{\alpha} _{k}d\alpha _{k}}{2\pi i}$. Here, $\bar{\boldsymbol{\alpha}}$ denotes the complex conjugate of $\boldsymbol{\alpha}$. The propagating function $\mathcal{J}(\boldsymbol{\bar{\alpha}}_f, \boldsymbol{\alpha}^{\prime}_f; t|\boldsymbol{\bar{\alpha}}_i, \boldsymbol{\alpha}^{\prime}_i; 0)$ is expressed as the path integral governed by an effective action consisting of the free actions of the forward and backward
propagators of the system and the influence functional obtained from the integration of reservoir degrees of freedom.  After
evaluating the path integral, we get
\begin{eqnarray}
& & \mathcal{J}(\boldsymbol{\bar{\alpha}}_{f},\boldsymbol{\alpha }_{f}^{\prime};t|\boldsymbol{\bar{\alpha}}_{i},\boldsymbol{\alpha }
_{i}^{\prime };0)=\exp \Big\{\sum_{k=1,2}\big[u_k(t)\bar{\alpha}_{kf}\alpha _{ki}  \notag \\
&&~~~~~+\bar{u}_k(t)\bar{\alpha}_{ki}^{\prime }\alpha _{kf}^{\prime
}+[1-\left\vert u_k(t)\right\vert ^2]\bar{\alpha}_{ki}^{\prime }\alpha _{ki}\big]\Big\},  \label{prord}
\end{eqnarray}
where $u_k(t)$ satisfies
\begin{equation}\label{ut}
\dot{u}_k(t) + i\omega_k u_k(t) + \int^{t}_0 f_k(t - \tau)u_k(\tau)d\tau = 0
\end{equation}
with $u_k(0)=1$ and $f_k(x) \equiv \int J_k(\omega)e^{-i\omega x}d\omega$ under the continuous limit of the environmental modes. Combining with Eq. (\ref{prord}), the time-dependent state can be obtained from any initial state by evaluating the integration in Eq. (\ref{rout}). The exact decoherence dynamics, determined by Eq. (\ref{ut}), essentially depends on the so-called spectral density $J_k(\omega)\equiv\sum_{l}\left\vert g_{kl}\right\vert^{2}\delta (\omega -\omega _{k})$, which characterizes the coupling strength of the different environmental modes to the system with respect to their frequencies. In the continuum limit, it takes the form $J_k(\omega )=\eta_k \omega \big( \frac{\omega }{\omega _{c}}\big)^{n-1} e^{-\frac{\omega }{\omega_{c}}}$, where $\omega_{c}$ is a cutoff frequency, and $\eta_k $ is a dimensionless coupling constant. The environment is classified as Ohmic if $n = 1$, sub-Ohmic if $0 < n < 1$, and super-Ohmic for $n > 1$ \cite{Leggett1987}.  Different spectral densities manifest different non-Markovian decoherence dynamics.

To compare with the conventional Born-Markovian approximate description to such system, a master equation can be derived by taking the time derivative to Eq. (\ref{rout})
\begin{eqnarray}
\dot{\rho}(t) &=&\sum_{k=1,2}\{-i\Omega_k (t)[\hat{a}_{k}^{\dag }\hat{a}_{k},\rho (t)]+\Gamma_k
(t)[2\hat{a}_{k}\rho (t)\hat{a}_{k}^{\dag }  \notag \\
&&-\hat{a}_{k}^{\dag }\hat{a}_{k}\rho (t)-\rho (t)\hat{a}_{k}^{\dag }\hat{a}_{k}] \},  \label{mas}
\end{eqnarray}
where $\Gamma_k (t)+i\Omega_k (t)\equiv-\dot{u}_k(t)/u_k(t)$. It can be seen that Eq. (\ref{mas}) keeps the Lindblad form but with time-dependent shifted frequency $\Omega_k(t)$ and decay rate $\Gamma_k(t)$. All the backactions induced by the non-Markovian effect have been incorporated into these time-dependent coefficients self-consistently.

\section{Dynamical frozen of Gaussian QD}
Consider explicitly the initial state of the system as the two-mode squeezed state $|\psi(0)\rangle = \exp[r(\hat{a}_1\hat{a}_2 - \hat{a}^{\dag}_1\hat{a}^{\dag}_2)]|00\rangle$ with $r$ being the squeezing parameter. The time evolution of such state under Eq. (\ref{rout}) keeps the Gaussianity. The Gaussian state can be fully characterized by the covariance matrix $\sigma_{12} =
 \begin{pmatrix}
  \alpha_1 & \gamma \\
  \gamma^{T} & \alpha_2
 \end{pmatrix}$,
where $\alpha_k$ are the $2 \times 2$ covariance matrices for the $k$-th subsystems, and $\gamma$ is the matrix containing
the correlations between $(x_1,p_1)$ and $(x_2,p_2)$ with $\hat{x}_k={\hat{a}_k+\hat{a}_k^\dag\over \sqrt{2}}$ and $\hat{p}_k={\hat{a}_k-\hat{a}_k^\dag\over \sqrt{2}i}$. $\sigma_{12}$ can be easily estimated experimentally from the homodyne measurements to the amplitude quadratures $\hat{x}_k$ and $\hat{p}_k$. The QD for the Gaussian state can be calculated as follows. The total correlation for a bipartite system is given by the mutual information $\mathcal{I}(\rho)=S(\rho_1)+S(\rho_2)-S(\rho)$, where $S$ is the von Neumann entropy and $\rho_{1(2)}$ is the reduced density matrix of the 1 (2) subsystem. Another measure of mutual information that only quantifies the amount of classical correlations extractable by a Gaussian measurement is $\mathcal{C}_1(\rho)=S(\rho_1)-\inf_{\sigma_M}S(\rho_{1|\sigma_M})$, where $\sigma_M$ is the covariance matrix of the measurement on mode 2. As it only captures the classical correlations, the difference, $\mathcal{D}_1=\mathcal{I}(\rho)-\mathcal{C}_{1}(\rho)$, is a measure of Gaussian quantum correlation that is coined Gaussian QD. An explicit expression for this QD has been found~\cite{Adesso2010}:
\begin{equation}
\mathcal{D}(\sigma_{12})=\mathfrak{f}(\sqrt{I_2})-\mathfrak{f}(\nu_-)-\mathfrak{f}(\nu_+)+\mathfrak{f}(\sqrt{m } )
\label{discord}
\end{equation}
with $\mathfrak{f}(x)=(\frac{x+1}{2})\ln{\frac{x+1}{2}}-(\frac{x-1}{2})\ln{\frac{x-1}{2}}$ and
\begin{align}
m=\left\{
\begin{array}{l l l}
\frac{2 I_3^2+(I_2-1)(I_4-I_1)+2|I_3|\sqrt{I_3^2+(I_2-1)(I_4-I_1)}}{(I_2-1)^2}  \\
\frac{I_1 I_2-I_3^2+I_4-\sqrt{I_3^4+(I_4-I_1 I_2)^2-2C^2(I_4+I_1 I_2)}}{2 I_2}
\end{array}\right.,
\label{ddf}\end{align}
where the top refraction of Eq. (\ref{ddf}) applies if $(I_4-I_1 I_2)^2\leq I_3^2(I_2+1)(I_1+I_4)$, and the bottom fraction of Eq. (\ref{ddf}) applies otherwise. Here $I_k=\det{\alpha_k}$, $I_3=\det{\gamma}$, $I_4=\det{\sigma_{12}}$ are the symplectic invariants and $\nu_\pm^2=\frac{1}{2}(\delta\pm\sqrt{\delta^2-4I_4})$ with $\delta=I_1+I_2+2I_3$ are the symplectic eigenvalues. The explicit form of the time evolution of the two-mode squeezed state and its corresponding covariance matrix are given in Appendix \ref{cova}. With the obtained covariance matrix (\ref{coaf}), the Gaussian QD can be evaluated straightforwardly.

\begin{figure}[tbp]
  \centering
  \includegraphics[width=0.49\columnwidth]{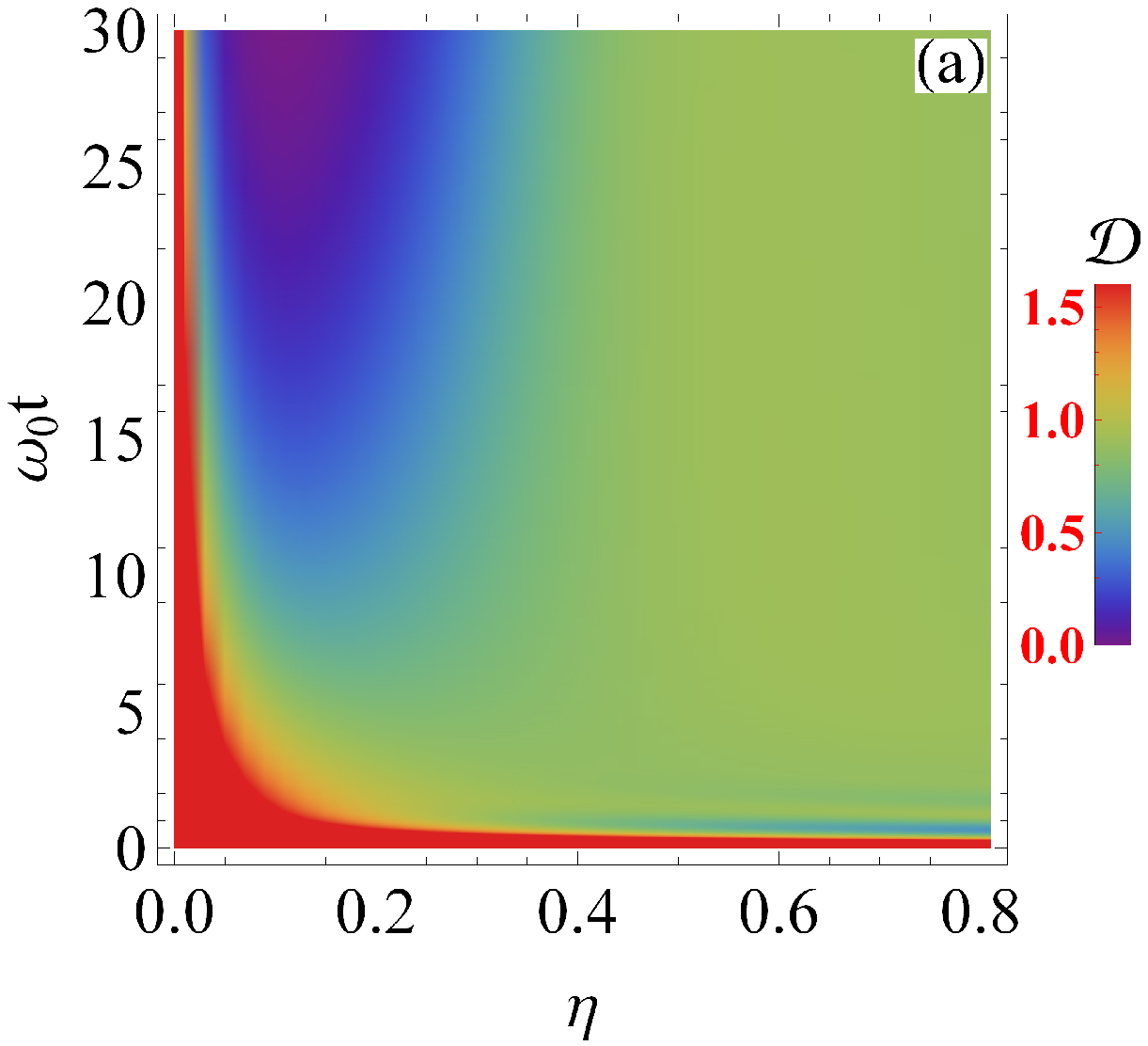}~\includegraphics[width=0.49\columnwidth]{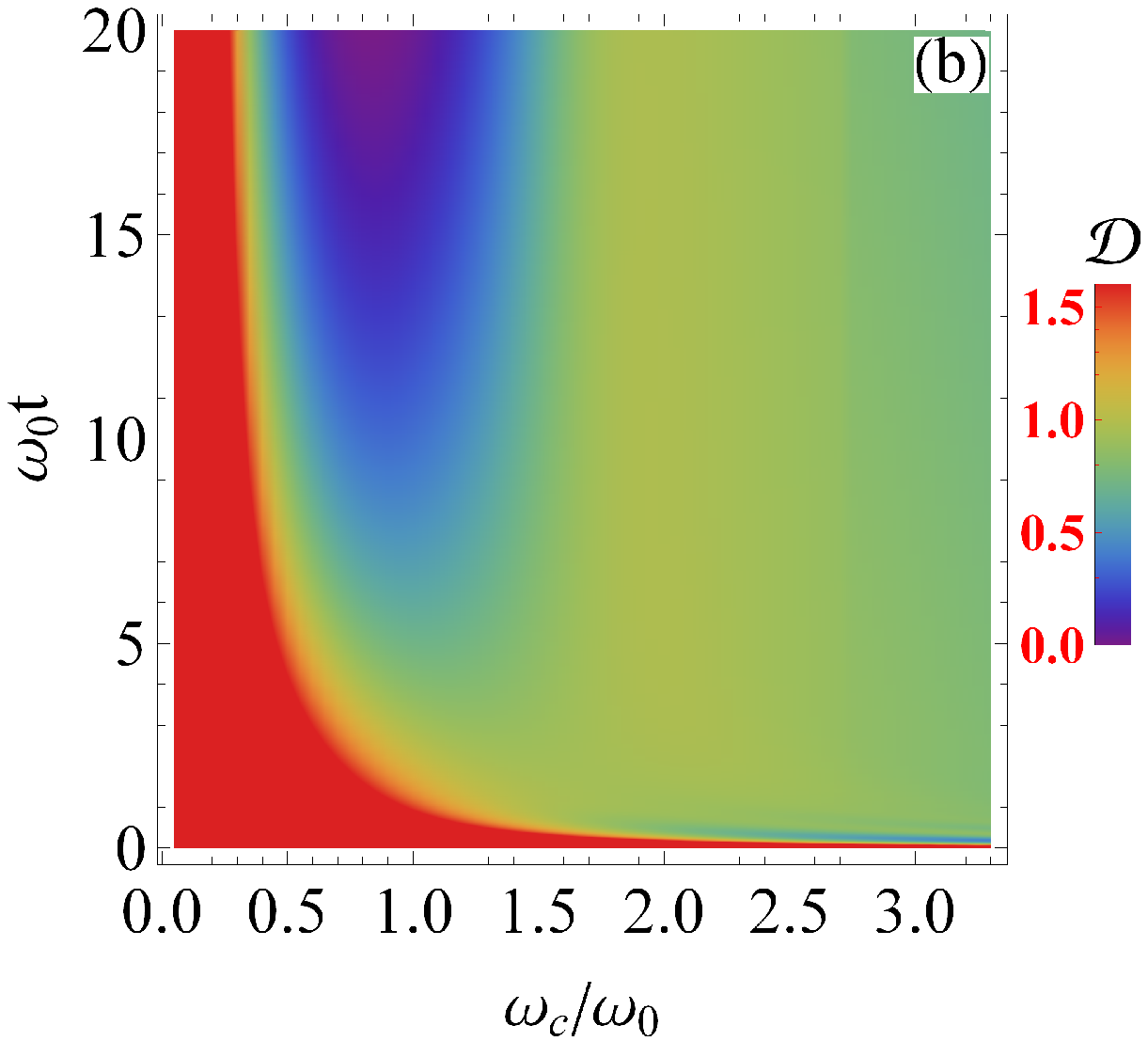}
  \caption{(Color online) The density plot of Gaussian QD vs $t$  for super-Ohmic spectral density in different (a) $\eta$ and (b) $\omega_c$. $r=1.0$ and $\omega_c/\omega_0=1.0$ in (a), and $\eta=0.08$ in (b).}\label{qdd}
\end{figure}
Choosing the super-Ohmic spectral density, explicitly $n=3$, as an example, we plot in Fig. \ref{qdd} the evolution of Gaussian QD for the initial two-mode squeezed state. It has been shown that the super-Ohmic spectral density can describe the phonon bath in one or three dimensions, depending on the symmetry properties of the strain field \cite{Weiss} and a charged particle coupled to its own electromagnetic field \cite{Barone}. Compared with the Ohmic and sub-Ohmic spectral densities, the super-Ohmic one is higher-frequency dominate, which will cause a strong modification to the short-time decoherence dynamics of the system. We can see from Fig. \ref{qdd}(a) that the Gaussian QD decays to zero and a larger $\eta$ induces a faster decay, which are qualitatively consistent with the results under the Markovian approximation, only when the coupling is vanishingly weak. With the increase of $\eta$, it is remarkable to find that the decay of the Gaussian QD tends to slow down even to be totally stabilized. This is dramatically contrary to one's expectation that a stronger coupling between the system and the reservoir always induces a more severe decoherence to the system. The similar frozen Gaussian QD can also be achieved with the increase of the cutoff frequency in Fig. \ref{qdd}(b).

We argue that the formation of a localized mode between each of the harmonic oscillators and its local reservoir plays an essential role in this frozen QD. To verify this, we perform a Fourier transform to Eq. (\ref{ut}) and obtain
\begin{equation}
y(E)\equiv\omega_0-\int_0^\infty {J(\omega)\over \omega-E}d\omega=E.\label{yee}
\end{equation}
One can see that $y(E)$ is a monotonically decreasing function in the region $E\in (-\infty,0)$. It means that Eq. (\ref{yee}) may have one and only one negative root if the system parameters fulfill $y(0)<0$. On the other hand, no further discrete root exists in the region $(0,+\infty)$ because that would make the integration in $y(E)$ divergent. After the inverse Fourier transform, the obtained $u_k(t)$ contributed from this discrete negative root will have a vanishing decay rate $\Gamma_k(t)$. This vanishing decay rate causes the decoherence inhabited in the system. It means that the discrete negative root for Eq. (\ref{yee}) actually corresponds to a stationary state to Eq. (\ref{ut}), which preserves the quantum coherence in its superposed components during time evolution. We call this stationary state the localized mode of the whole system \cite{Zhang2012}. For our super-Ohmic spectral density, we can readily show that the localized mode is formed when $\omega_0-2\eta\frac{\omega_c^3}{\omega_0^2}<0$ is fulfilled. This criterion gives a basic judgment on the condition under which the frozen Gaussian QD is present.

\begin{figure}[tbp]
  \centering
  \includegraphics[height=0.47\columnwidth]{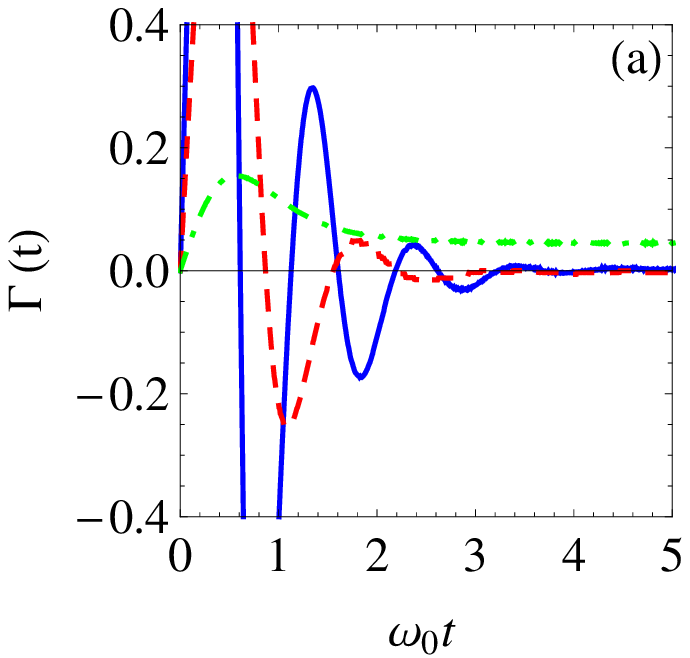}~\includegraphics[height=0.47\columnwidth]{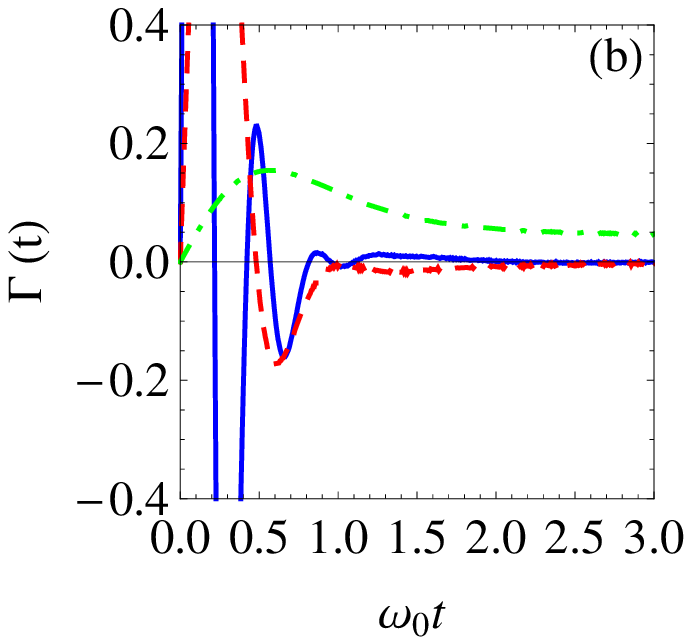}
  \caption{(Color online) The decay rate when (a) $\omega_c/\omega_0=1.0$ and $\eta=0.08$ (dot-dashed green line), $0.5$ (dashed red line), and $1.0$ (solid blue line) and (b) $\eta=0.08$ and $\omega_c/\omega_0=1.0$ (dot-dashed green line), $2.0$ (dashed red line), and $3.0$ (solid blue line). The localized mode is formed when (a) $\eta>0.5$ and (b) $\omega_c>1.84\omega_0$.  }\label{ug}
\end{figure}

To verify the dynamical consequence of the formed localized mode, we plot in Fig. \ref{ug} the decay rate in the case of Figs. \ref{qdd}(a) and \ref{qdd}(b). We can see that if the localized mode is absent, the decay rate stays positive and tends to a positive value, which, as expected, will induces monotonic decoherence to the system, as shown in Fig. \ref{qdd}(a) when $\eta<0.5$ and in Fig. \ref{qdd}(b) when $\omega_c<1.84\omega_0$. On the contrary, if the localized mode is present, the decay rate is transiently negative, which manifests that the lost information/energy of the system returns back from the reservoir. Another characteristic that is different from the case when the localized mode is absent is that the decay rate tends to zero asymptotically. This vanishing decay rate causes the decoherence of the system to cease in the long-time limit. This gives an explanation why a strong coupling can induce a suppressed decoherence in Fig. \ref{qdd}. Such anomalous decoherence also manifests as the deviation from the exponential decay of $|u(t)|^2$ under the Born-Markovian approximation, as shown in Appendix \ref{anomde}.

From the above analysis, we can conclude that the frozen Gaussian QD is present due to an interplay between the formed localized mode and the non-Markovian effect. The localized mode provides an ability to freeze the Gaussian QD, while the non-Markovian effect provides a dynamical way to freeze the Gaussian QD. The mechanism of the stable Gaussian QD frozen in our system is linked to the non-Markovian memory effect of the harmonic oscillator with its local reservoir when the localized mode is formed. It is much different from the case of two harmonic oscillators coupled to a common reservoir \cite{Paz2012,Correa2012}, where a stable QD is established due to an indirect interaction between the two harmonic oscillators induced effectively by the common reservoir.

It is noted that our result is also of benefit to the analysis of entanglement under the same decoherence setting. Since the decoherence is suppressed when the localized mode is formed, we also could expect a finite entanglement preservation in the steady state. In this case, the Gaussian QD shows no qualitative difference from the entanglement. However, in the parameter regime where the localized mode is absent, it can be confirmed that the entanglement always decays to zero more rapidly than the Gaussian QD does. This is consistent with the previous result that the QD is more robust than the entanglement to local decoherence \cite{Werlang2009,Fanchini2010,Wang2010,Ferraro2010,Ge2010,Xu2010,Madsen2012,Vasile2010,Isar2012}.

\section{Physical realization}
\begin{figure}[tbp]
  \centering
  \includegraphics[width=0.75\columnwidth]{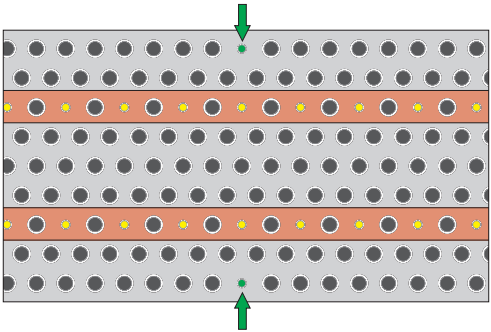}\\
  \caption{(Color online) Two initially correlated cavity fields propagating in two cavity arrays formed in a photonic crystal.   }\label{cpa}
\end{figure}
With the basic criterion at hand, we can see that the frozen Gaussian QD upon which we elaborated is a generic phenomenon in open quantum systems, irrespective of the form of the spectral density. A best candidates with which to observe our prediction is the system of two arrays of coupled cavities, which can now be realized experimentally in micro-disc cavities coupled by one tapered optical fiber \cite{Barclay2006}, in a photonic crystal system \cite{Hennessy2007, Notomi2008, Majumdar2012}, and synthesized in an optical waveguide array system \cite{Verbin2013,Rechtsman2012}. In Fig. \ref{cpa}, we depict the schematic illustration to this scheme realized in a photonic crystal system. Here, two initially correlated quantized optical fields are fed into the two system cavities. With some probability the optical fields in the two system cavities will hop, respectively, to the two spatially separated coupled cavity arrays. Each of the local systems is governed by $\hat{H}^{(1)}=\omega_0\hat{a}^\dag\hat{a}+\omega_C\sum_{j=0}^{N-1}\hat{ b}^\dag_j \hat {b}_j+(g\hat{a}^\dag \hat{b}_0+\xi\sum_{j=0}^{N-2}\hat{b}_{j+1}^\dag\hat{b}_j+\text{h.c.})$.
A Fourier transform $\hat{b}_j=\sum_k\hat{b}_ke^{ikjx_0}$ recasts $\hat{H}^{(1)}$ into \begin{equation}\hat{H}^{(1)}=\omega_0\hat{a}^\dag\hat{a}+\sum_{k}\epsilon_k\hat{ b}^\dag_k \hat {b}_k+{g\over\sqrt{N}}\sum_k(\hat{a}^\dag\hat{b}_k+\text{h.c.})\end{equation} with $\epsilon_k=\omega_C+2\xi\cos kx_0$, and $x_0$ being the spatial separation between the two neighbor cavities of the cavity arrays. One can notice that the dispersion relation of the field in such structured reservoirs shows finite band width, which can induce a strong non-Markovian even in the weak and intermediate coupling regimes.

\begin{figure}[tbp]
  \centering
  \includegraphics[width=0.48\columnwidth]{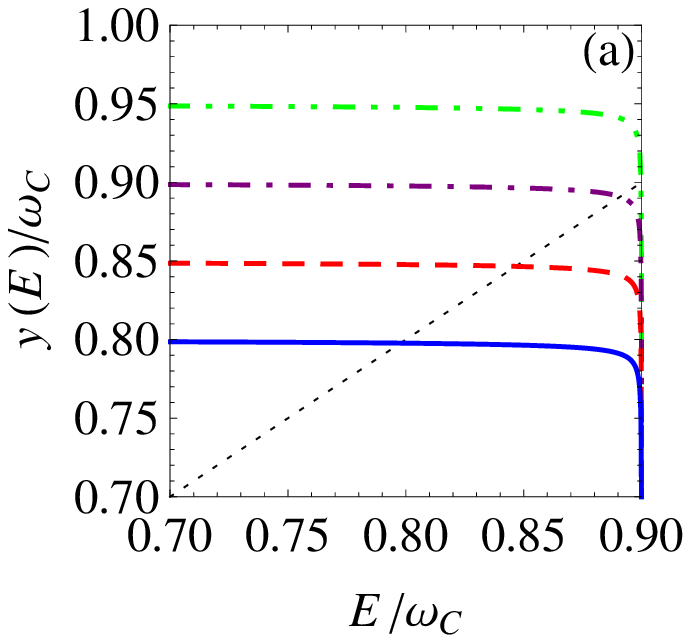}~\includegraphics[width=0.51\columnwidth]{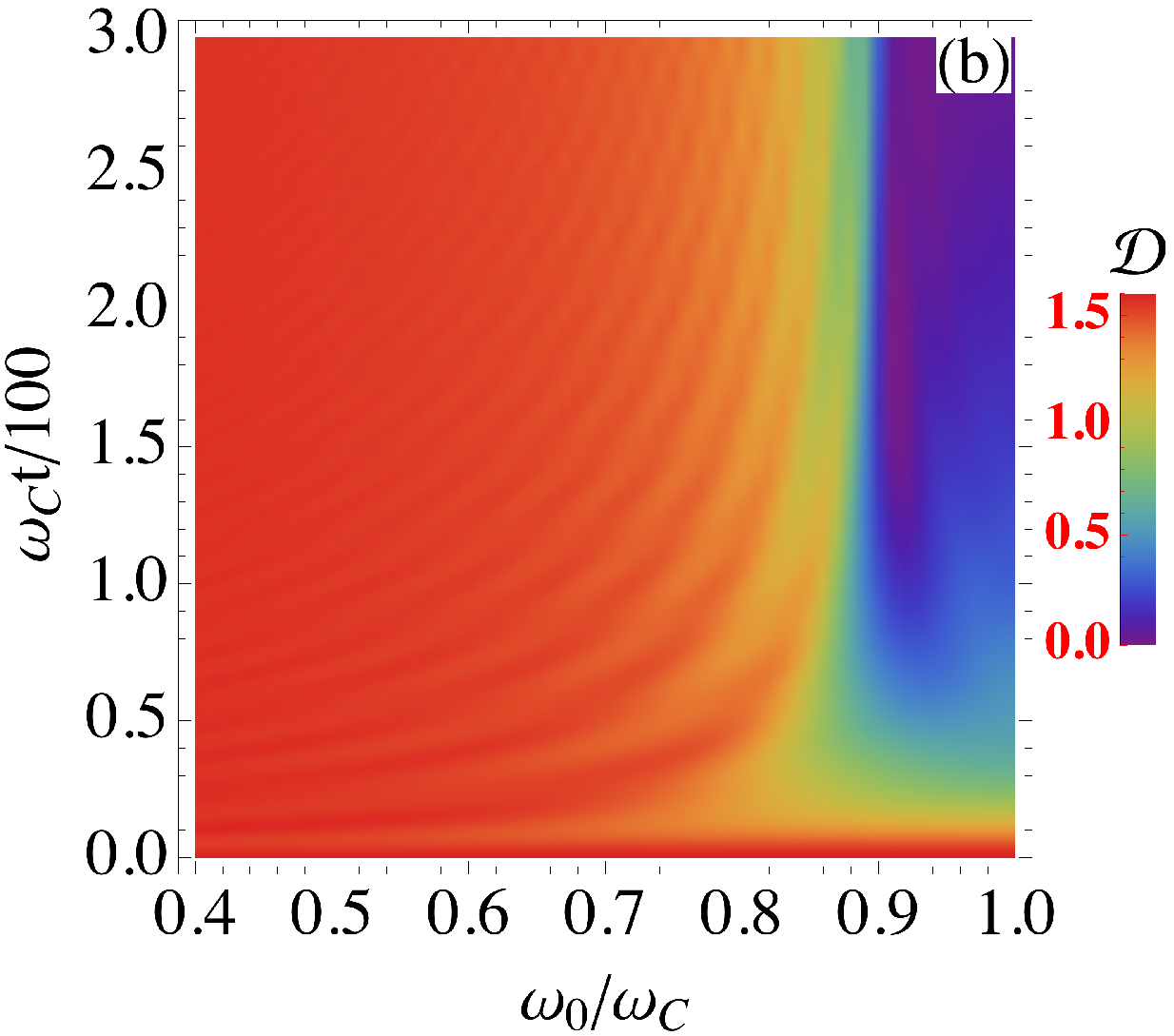}
  \caption{(Color online) (a) The formation of a localized mode manifested by the intersection point of the dotted line with the lines when $\omega_0=0.95\omega_C$ (dot-dot-dashed green line), $0.9\omega_C$ (dot-dashed purple line), $0.85\omega_C$ (dashed red line), and $0.8\omega_C$ (solid blue line).  (b): The density plot of Gaussian QD vs $t$ in different $\omega_0$. $\xi=0.05\omega_C$, $g=0.02\omega_C$, and $N=200$ have been used. }\label{cwa}
\end{figure}

In Fig. \ref{cwa}(a), we plot the possible formation of the localized mode manifested by the intersection points between the dotted line and each line in different parameter regimes. It can be seen that if there is no intersection point, which means the localized mode is absent, then the Gaussian QD, as shown in Fig. \ref{cwa}(b), decays to zero. Whenever the localized mode is formed, certain finite Gaussian QD can be frozen in the steady state. As an interesting observation, we find that the frozen Gaussian QD in this case is even as large as its initial value. It means that the detrimental effect from decoherence is almost eliminated. Another interesting observation is that the frozen QD can be obtained even if there is no strong coupling between the system and the reservoirs. This reduces greatly the experimental difficulty in the practice.

\section{Conclusions}  We have revealed a mechanism under which the decoherence of QD can be avoided and a finite QD can be frozen in the steady state. The underlying physics is the interplay between the formed localized mode and the non-Markovian effect. We have also proposed an experimentally accessible scheme to observe our prediction in a coupled cavity array system realized in a photonic crystal platform \cite{Hennessy2007, Notomi2008, Majumdar2012}. Our result suggests the controllability of decoherence by reservoir engineering \cite{Barreiro2011,Murch2012}. Our finding provides significant progress in the practical continuous-variable quantum information processing.

\section*{Acknowledgement}
This work is supported by the Fundamental Research Funds for the Central
Universities, by the NSF of China (Grants No. 11175072 and No. 11174115), and by the National Research Foundation and Ministry of Education,
Singapore (Grant No. WBS: R-710-000-008-271).

\appendix
\section{The covariance matrix}\label{cova}
The initial state can be represented in the coherent-state representation as
\begin{equation}\label{dfds}
\rho(\boldsymbol{\bar{\alpha}}_i, \boldsymbol{\alpha}^{\prime}_i; 0) =
\frac{\exp[-\tanh r(\bar{\alpha}_{1i}\bar{\alpha}_{2i} + \alpha^{\prime}_{1i}\alpha^{\prime}_{2i})]}{\cosh^2r}.
\end{equation}Substituting Eq. (\ref{dfds}) into Eq. (\ref{rout}), we can obtain the evolved state as
\begin{equation}
\rho (\boldsymbol{\bar{\alpha}}_{f},\boldsymbol{\alpha }_{f}^{\prime
};t)=a\exp [\sum_{k\neq k^{\prime }}(\frac{b}{2}\bar{\alpha}_{kf}\bar{\alpha}%
_{k^{\prime }f}+c\bar{\alpha}_{kf}\alpha _{kf}^{\prime }+\frac{b^{\ast }}{2}%
\alpha _{kf}^{\prime }\alpha _{k^{\prime }f}^{\prime })],  \label{final}
\end{equation}where
\begin{eqnarray}
a & = & \frac{1}{\cosh ^{2}\left\vert r\right\vert [1-\tanh ^{2}\left\vert r\right\vert (1-\left\vert u(t)\right\vert ^{2})^{2}]}, \\
b & = & \frac{-\tanh \left\vert r\right\vert u(t)^{2}}{1-\tanh ^{2}\left\vert r\right\vert (1-\left\vert u(t)\right\vert ^{2})^{2}}, \\
c & = & \frac{\tanh ^{2}\left\vert r\right\vert (1-\left\vert u(t)\right\vert ^{2})\left\vert u(t)\right\vert ^{2}}{1-\tanh ^{2}\left\vert r\right\vert
(1-\left\vert u(t)\right\vert ^{2})^{2}}.
\end{eqnarray}

For the continuous-variable (Gaussian-type) bipartite state, its density matrix is characterized by the covariance matrix defined as the second moments of the quadrature vector $\hat{X} = (\hat{x}_1, \hat{p}_1, \hat{x}_2, \hat{p}_2)$,
\begin{equation}
\sigma_{ij} = \langle\Delta \hat{X}_i\Delta \hat{X}_j + \Delta \hat{X}_j\Delta \hat{X}_i\rangle,
\end{equation}
where $\Delta \hat{X}_i = \hat{X}_i - \langle \hat{X}_i\rangle$, and $\hat{x}_i = \frac{\hat{a}_i +\hat{ a}^{\dag}_i}{\sqrt{2}}$, $\hat{p}_i = \frac{\hat{a}_i - \hat{a}^{\dag}_i}{i\sqrt{2}}$. From the time-dependent state (\ref{final}), the covariance matrix for the harmonic oscillators
can be calculated straightforwardly,
\begin{equation}
\sigma=2\left(
\begin{array}{cccc}
\frac{y(1+d)}{2(1-d)^{2}} & 0 & \frac{a\text{Re}[b]}{x} & \frac{a\text{Im}[b]%
}{x} \\
0 & \frac{y(1+d)}{2(1-d)^{2}} & \frac{a\text{Im}[b]}{x} & \frac{-a\text{Re}%
[b]}{x} \\
\frac{a\text{Re}[b]}{x} & \frac{a\text{Im}[b]}{x} &
\frac{y(1+d)}{2(1-d)^{2}}
& 0 \\
\frac{a\text{Im}[b]}{x} & \frac{-a\text{Re}[b]}{x} & 0 & \frac{y(1+d)}{%
2(1-d)^{2}}%
\end{array}%
\right) ,\label{coaf}
\end{equation}%
where $x=[(1-c)^{2}-\left\vert b\right\vert ^{2}]^{2}$, $y=\frac{a}{1-c}$,
and $d=c+\frac{\left\vert b\right\vert ^{2}}{1-c}$.

\section{Anomalous decoherence}\label{anomde}

\begin{figure}[t]
  \centering
  \includegraphics[width=0.49\columnwidth]{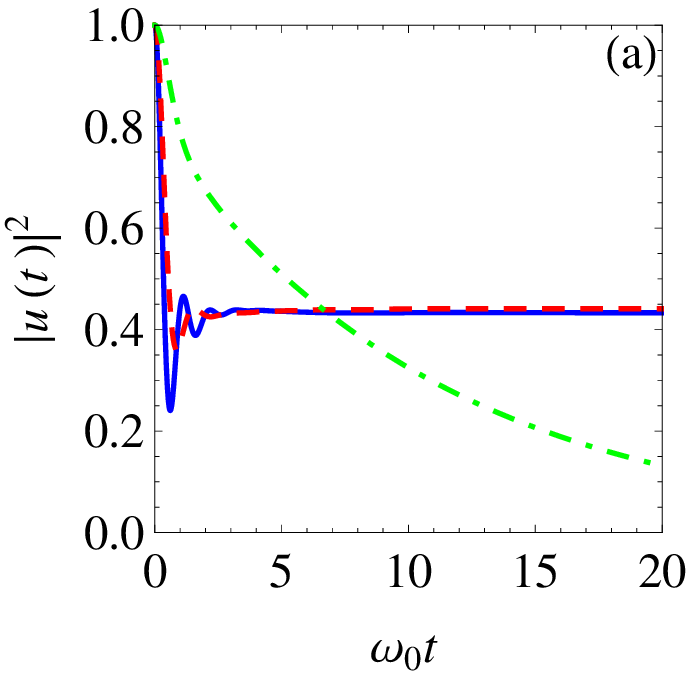}~\includegraphics[width=0.49\columnwidth]{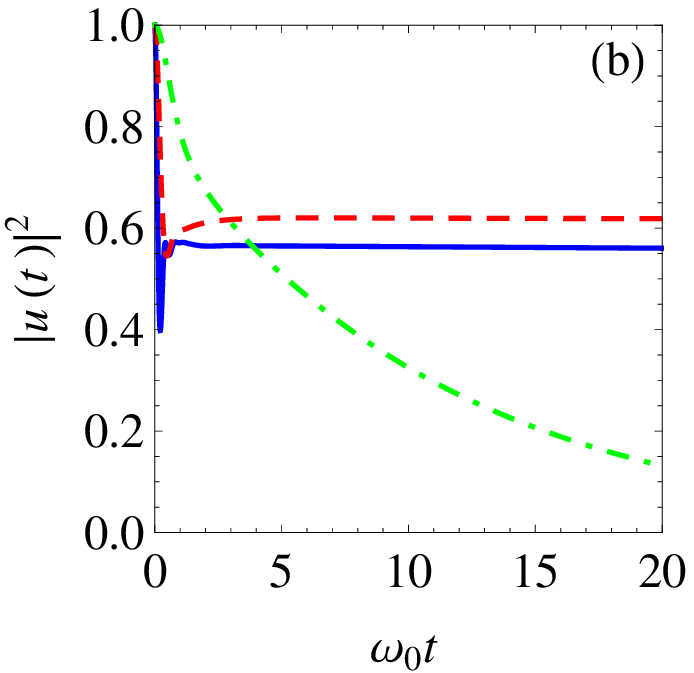}
  \caption{(Color online) The corresponding $|u(t)|^2$ of Figs. \ref{ug}(a) and \ref{ug}(b). The localized mode is formed when (a) $\eta>0.5$ and (b) $\omega_c>1.84\omega_0$.  }\label{ugt}
\end{figure}

Accompanying the formation of the localized mode of the whole system, the dynamics of the reduced system is inhibited. This can be verified by the time-dependent behaviors of $u(t)$. In Fig. \ref{ugt}, we plot the evolution of $|u(t)|^2$ corresponding to the parameter regimes used in Fig. \ref{ug}(a) and \ref{ug}(b), respectively. We can see that with the formation of the localized mode above the critical point $\eta=0.5$ for Fig. \ref{ugt}(a) and $\omega_c=1.84\omega_0$ for Fig. \ref{ugt}(b), the time-dependent behavior of $|u(t)|^2$ shows qualitative changes. If the localized mode is absent, $|u(t)|^2$ decays to zero monotonically, which is consistent with the results under the Born-Markovian approximation. On the other hand, if the localized mode is present, $|u(t)|^2$ tends to a finite value after transient oscillation. It indicates the ceasing of the decoherence in the long-time limit, which is also consistent with the vanishing decay rate in Fig. \ref{ug}. It deviates qualitatively from the results under Born-Markovian approximation. This shows clearly that the non-Markovian effect can induce not only transient oscillation, but also dramatic change on the steady state behavior to the open quantum system. Equipped with this anomalous decoherence, it is not hard to understand the frozen Gaussian QD revealed in our work.

\end{document}